\documentclass[epj]{svjour}

\usepackage{latexsym}
\usepackage{graphics}

\newcommand{\be}{\begin{equation}}

\newcommand{\ee}{\end{equation}}

\begin{document}


\title{Scissors modes in crystals with cubic symmetry}




\author{Keisuke Hatada\inst{1,2} \and Kuniko Hayakawa\inst{2, 3} \and Fabrizio Palumbo\inst{2}}



%

\institute{Instituto de Ciencia de Materiales de Arag\'on, CSIC-Universidad de Zaragoza, 50009 Zaragoza, Spain \and
INFN Laboratori Nazionali di Frascati, c.p. 13, I-00044 Frascati, Italy \and
Centro Fermi, Compendio Viminale, Roma I-00184, Italy}

\date{}

\abstract{We recently suggested that the Scissors Mode (a collective excitation in which one system rotates with respect another one conserving its shape)
can occur in crystals with axially symmetric atoms as a precession of these atoms around the anisotropy axis of their cells, giving rise to a form of dichroism. In the
present paper we investigate how the Scissors Mode can be realized in crystals with cubic symmetry and evaluate its
photo-absorption cross-section. This turns out to be of the same order of magnitude as that for crystals with axially symmetric atoms, but does not exhibit 
any correlation between the direction of the photon and the axes of the cell.
}


\maketitle

\section{Introduction}

The Scissors Mode is a collective excitation in which two systems move with respect to each other conserving their shape.
It was first predicted to occur in deformed atomic nuclei~\cite{loiu} by a semiclassical calculation in which protons and
neutrons were assumed to form two distinct bodies to be identified with scissors blades. Their relative rotation
generates a magnetic dipole moment whose coupling with the electromagnetic field provides the signature of the mode.

After its discovery~\cite{bohle} in a rare earth nucleus, $^{156}$Gd, and its systematic
experimental and theoretical investigation~\cite{Ende} in all deformed atomic nuclei, it was predicted 
to occur in several other systems including metal clusters~\cite{Lipp}, quantum dots~\cite{Serr},
Bose-Einstein~\cite{Guer} and Fermi~\cite{Ming} condensates and crystals~\cite{Hata}. 
In all these other systems one of the blades of the scissors must be identified
with a moving cloud of particles (electrons in metal clusters and
quantum dots, atoms in Bose-Einstein and Fermi condensates, an atom in a cell in crystals) and the other one with a structure at rest (the trap in Bose-Einstein and Fermi condensates,
the lattice in metal clusters, quantum dots and crystals). Among these other systems until now the existence of the Scissors mode has been confirmed only in Bose-Einstein condensates\cite{Mara}.

The case of crystals has been studied only for atoms with axial symmetry and ionic bond, in a semiclassical model ~\cite{Hata} in which an atom 
whose charge density profile is approximately ellipsoidal is regarded as a rigid body whose symmetry axis can precess around the axes of its cell under the electrostatic force generated by the ligands.
If the deformation is sufficiently large it can rotate only around the anisotropy axis of the cell because the restoring force is too strong to allow a precession around the other axes.
With such a geometry the Scissors Mode gives rise to a magnetic dichroism, with a strong M1 transition amplitude which is maximum
when the momentum of the absorbed photon is parallel to the anisotropy axis of the cell and zero when it is orthogonal to it. A numerical estimate was done for LaMnO$_3$. It was found that the energy of the Scissors Mode
is about 4 eV (but see Section 3) and the M1 transition amplitude of the order of 1.5 a.u.. For small deformations one should expect, in analogy with atomic nuclei~\cite{Palu}, 
a splitting of this collective mode corresponding to precessions around all the cell axes.

There is a great deal of compounds whose structure has cubic symmetry,
either 
at a global level (cubic space groups) or at a local level. We have in
mind 
the transition metal monoxides that crystallize in the rocksalt
structure, like  
TiO and NiO, other that adopt the rare stechiometry MO$_3$, like
ReO$_3$, all
the ternary compound having a perovskite cubic structure, named after
the mineral 
CaTiO$_3$, etc.., not to mention the many transition metal complexes in 
octrahedral coordination, like Fe(CN)$_6^{3-}$ or Co(NH$_3)_6^{3+}$.
Among the Rare Earth compounds we mention Europium oxide (EuO) as a
prototypical 
example of cubic crystal, but one can find other examples in cubic Laves
phases of 
Rare Earth and pnictides. Examples of Rare Earth ions in octahedral coordination
can be found also in ~\cite{McCl}.

Obviously the analysis of Ref.~\cite{Hata}
cannot be valid for such systems, but the magnetic nature of the Scissors Mode makes its investigation more interesting because 
magnetic anisotropy is at the origin of interesting technological applications of some such systems, including magnetic storage devices and sensors,
spin-torque nano-oscillators for high-speed spintronics and spin-optics~\cite{Smen}. We would like in fact to remind that in the cases studied so far,
the Scissors Modes provides specific pieces of information about the structure of the systems. In nuclear physics it is related to the superfluidity of deformed nuclei, in Bose-Einstein Condensates it
provides a signature of superfluidity, in metal clusters it is predicted to be responsible for paramagnetism and in crystals with axially symmetric atoms of a form of magnetic dichroism. In a separate work we will relate  Scissors Modes in crystals with cubic symmetry in the Rare Earth region to Spin-Orbit Locking, a conjectured atomic structure in which, due to a strong spin-orbit force, the charge profile of an atom in a cell is firmly bound to the atomic spin.
More generally  the study of the possible occurrence of such collective modes is interesting  in understanding to which extent it is a universal feature of finite many-body systems. 

 In the present paper we investigate in which way the Scissors mode can be realized in crystals with cubic symmetry. We find that if it
 is excited by a photon whose momentum is parallel to one of the symmetry axes, the motion resembles that of a balance-wheel but,
 unlike the case of atoms with axial symmetry, the absorption cross-section does not depend on the direction of the photon momentum.
 
 In Section 2 we describe the semiclassical model which we adopt and we evaluate excitation energy and photoabsorption cross-section. In Section
 3 we discuss the numerical estimates of the parameters of the model and their relation to some conceptual issues. We conclude in Section 4.

\section{The model}

We assume a frame of reference with the $x_1, x_2, x_3$-axes parallel to the axes of the cell, and we denote by $\xi_1,\xi_2, \xi_3$ the axes of the 
principal frame of inertia of the atom, regarded as a rigid body of cubic profile, as shown in Fig. 1. If we call $\phi_i$ the rotation angle around the $x_i$-axis, the Hamiltonian of the atom is
\begin{equation}
H=\sum_{i=1}^3 \left( - {   \hbar^2  \over 2 \, {\mathcal I}} \, {\partial^2 \over \partial \phi_i^2} + V ( \phi_i) \right)
\end{equation}
where ${\mathcal I}$ is the moment of inertia of the atom with respect to any of its axes. This expression is justified for the lowest lying excitations for which,
due to the cubic symmetry, we do not need to express the kinetic energy in terms of the components of the full angular momentum. Moreover, because of this symmetry the potential
must satisfy the relation $R_i({\pi \over 2}) V(\phi_i)=V(\phi_i)$, where $ R_i(\phi_i)$ is the rotation operator through the angle $\phi_i$ around the $i$-axis.

 \begin{figure}[htbp]
     \resizebox{70mm}{!}{\includegraphics{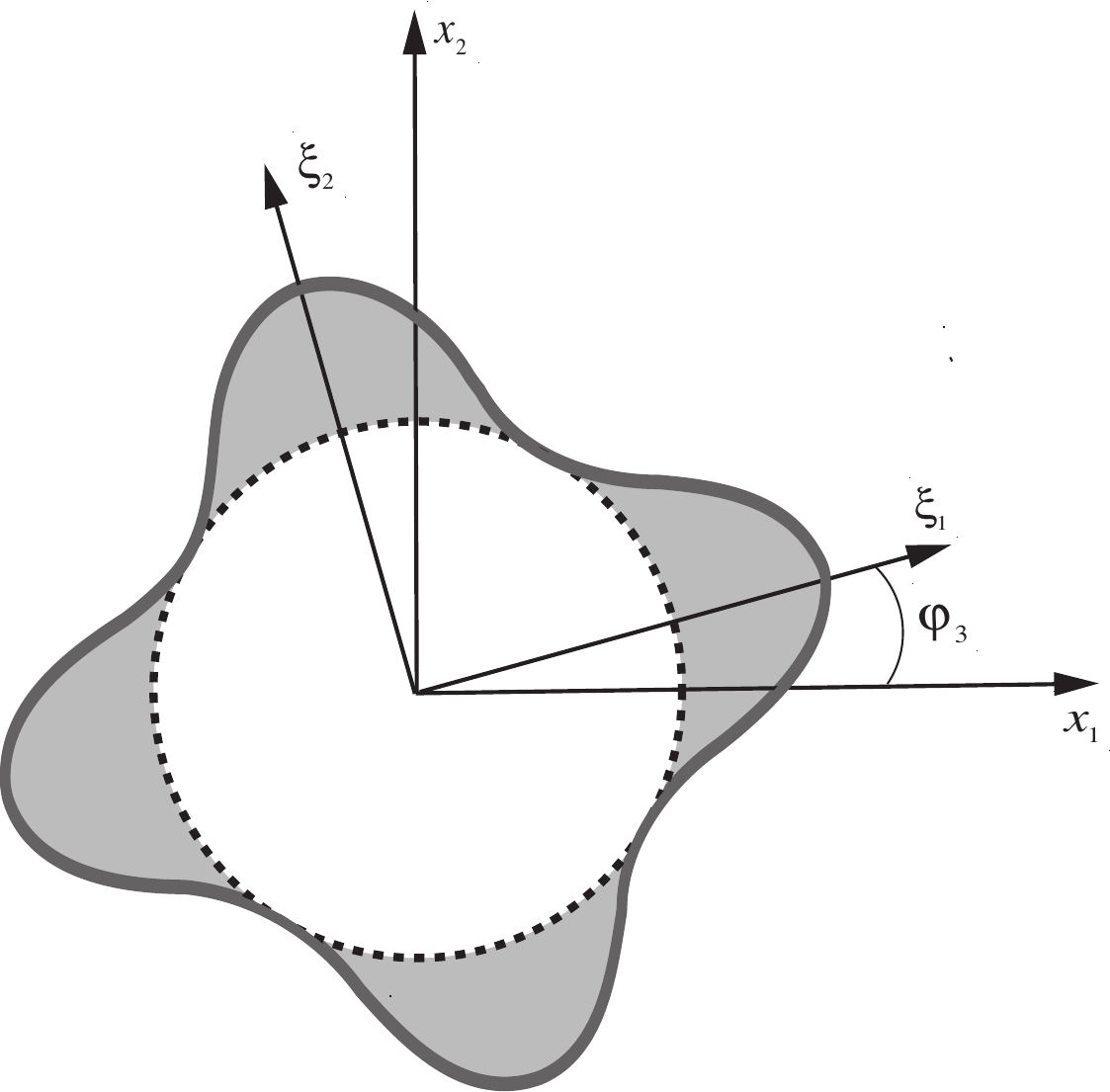}}
         \caption{A section of an atom with cubic symmetry in the plane $x_1, x_2$.  Its principal frame of inertia, with axes $\xi_1, \xi_2$, oscillates around the  $x_3 = \xi_3$  axis as a balance-wheel.
         Its profile can approximately be obtained by a superposition of ellipsoids. Only the charge external to the sphere contributes to both the restoring force constant and the moment of inertia.}
\end{figure}

Therefore it is sufficient to define it in the interval $- {\pi \over 4} \le \phi \le {\pi \over 4}$,
in which we assume the harmonic form 
\begin{eqnarray}
V(\phi)= { 1\over 2} C  \phi^2 \,,\,\,\,- {\pi \over 4} \le \phi \le {\pi \over 4} \,.
\end{eqnarray}
The eigenfunctions normalized in this interval for each of the angles $\phi_i$ are 
harmonic oscillator wave functions
\begin{equation}
\lambda_n(\phi)= { 1 \over  \sqrt{  \sqrt{\pi} 2^n n!  \, \phi_0 }} \, H_n \left( {\phi \over \phi_0}\right)
\exp\left( - { 1\over 2}  {\phi^2 \over \phi_0^2} \right) 
\end{equation}
where $H_n$ are Hermite polynomials and
\begin{equation}
\phi_0^2=
{ \hbar \over \sqrt{{\mathcal I} C}} \label{phi}
\end{equation}
with eigenvalues
\be
 E_n = \left( { 1\over 2} + n \right) \hbar \omega \,,  \,\,\,\,\omega = \sqrt{ {C\over {\mathcal I}}}\,.
\ee
Because of the cubic symmetry the eigenfunctions in the full range of $\phi$ are
\begin{eqnarray}
\psi_n(\phi_i) &=& { 1 \over 2} \left[ \lambda_n(\phi_i) + R_i\left( { 1 \over 2} \pi \right)\lambda_n(\phi_i) \right.
\nonumber\\
&+& \left.  R_i (\pi)\lambda_n(\phi_i) + R_i \left( {3 \over 2} \pi \right)\lambda_n(\phi_i) \right] \,.
\end{eqnarray}
We will consider only transitions to the lowest excited states, denoted by $\Psi_i, i=1,2,3$ if the oscillation takes place around the $i$-th axis, and we well denote by $\Psi_0$ the ground state. For instance
 \be
 \Psi_1(\phi_1, \phi_2, \phi_3) = \psi_1(\phi_1)\psi_0(\phi_2)\psi_0(\phi_3) \,.
 \ee
The interaction Hamiltonian with the electromagnetic field is
\be
H_I = \sum_l - {e \over m} {\bf p^{(l)}} \cdot  {\bf A}({\bf r}^{(l)}) + { 2 \mu \over \hbar } \, {\bf S^{(l)}} \cdot  \nabla^{(l)} \times {\bf A}({\bf r}^{(l)}) 
\ee
where $m, e, \mu, {\bf S}, {\bf p}$ and ${\bf r}$ are the mass, electric charge, magnetic moment, spin, momentum and position vector of the electron, and the sum extends over 
the electrons of the atom. Writing the vector potential as
\be
{\bf A} = \sum_{\lambda}  {\vec \varepsilon}^{(\lambda)} \exp (i {\bf k} \cdot {\bf r})
\ee
where $  {\vec \varepsilon}^{(\lambda)}$ is the photon polarization vector, the 
absorption cross-section for a photon of polarization $\lambda$ and energy $\hbar \omega = \hbar k \,c$, $c$ being the velocity of light is
\be
\sigma^{(\lambda)} = { 4 \pi^2 \alpha \hbar^2 \over m^2 \omega^2} \sum_i |M_i^{(\lambda)}|^2 
\ee
where $\alpha$ is the fine structure constant and 
\be
M_i^{(\lambda)} = \langle \Psi_i|e^{i {\bf k} \cdot {\bf r}} {\vec \varepsilon}^{(\lambda)} \cdot 
\sum_l \left( \nabla^{(l)}- i \, {\bf S}^{(l)} \times \nabla^{(l)}  \right) |\Psi_0 \rangle \,.  \label{M}
\ee
In our semiclassical approximation the deformed atoms rotate as rigid bodies. Therefore when they perform rotational oscillations around
the $i$-th axis the velocity of an electron at position $r, \phi$ is
\be 
{\bf v}=  {d\phi_i  \over dt}  \, {\bf r} \times  {\hat {\bf x}}_i
\ee
where ${ \hat {\bf x}}_i$ is the unit vector in the $i$-th direction. The time derivative of $\phi$ is determined by the equation of motion
 \be 
 {d \phi \over d t} =- i { 1\over \hbar} \,  [H, \phi] = i \, {\hbar \over  {\mathcal I}} {\partial  \over \partial \phi} 
 \ee
so that
\be 
\nabla = -  {m \over  {\mathcal I}}  {\bf r} \times  {\hat {\bf x}_i} \, {\partial  \over \partial \phi_i} \,.
\ee
 In the electric dipole approximation (neglecting the 
exponential ) the matrix element (\ref{M}) vanishes, because the contributions from the four regions of $\phi_i$ cancel out with each other.
Enforcing the cubic symmetry is essential to get this result, which corresponds to the physical fact that the electron velocities are opposite in opposite regions of ${\phi}$.
Taking into account the first term in the expansion of the exponential we get
\begin{eqnarray}
M_i^{(\lambda)} &=& - {2m \over {\mathcal I} } {\vec \varepsilon}^{(\lambda)} \cdot  \, \sum_l \langle \Psi_i|  \left( i {\bf k}\times {\hat {\bf x}}_i 
- {\bf k} \cdot {\bf S}^{(l)} \, {\hat {\bf x}}_i \right) 
\nonumber\\
& & \times \, (r^{(l)})^2 { \partial \over \partial \phi_i}
 |\Psi_0\rangle \,.
\end{eqnarray}
Now we make the approximation
\be
\sum_l  m \, \langle( r^{(l)} )^2 \rangle   \approx { 3 \over 2} {\mathcal I}\,.
\ee
Concerning the sum over the spins we  observe that in some of the compounds with cubic symmetry the spins of localized electrons
are not balanced. The atoms of these  compounds  have a net spin of the order of several units of electronic spin. If the spin-orbit
force is strong with respect to the crystalline field, the atomic spin is firmly coupled to the charge density profile \cite{Gerr}, a coupling which we
call Spin-orbit Locking. In this case the atomic spin will follow the rotation of the charge profile.
Since the rotation angle is very small (see below), we can  make the approximation
 \be
 \sum_l {\bf S}^{(l)} \approx  { 1 \over 2} \eta \,{\hat {\bf S}}
 \ee
where $\eta$ is the ratio of the number 
of electrons which contribute to the spin over the number of electrons which participate to the collective motion and $ {\hat {\bf S}}$
is the unit vector in the direction of the spin. At last
 \be
M_i^{(\lambda)} ={ 3 \over 4 {\sqrt 2} \, \phi_0} {\vec \varepsilon}^{(\lambda)} \cdot  \left( {\bf k}\times {\hat {\bf x}}_i 
- { 1\over 2} i \, \eta \, {\bf k} \cdot {\hat {\bf S}} \, {\hat {\bf x}}_i \right) \,. 
\ee
Therefore
\be
\sum_{i=1}^3  \sum_{\lambda} | M_i^{(\lambda)} |^2 = { 9 \over 16 \, \phi_0^2}  \left( k^2 + { 1\over 4} \eta^2  ({\bf k} \cdot {\hat {\bf S}} )^2  \right)\,.
\ee
We see that the contribution of the spin is negligible, so that study of excitation of the Scissors Modes by photoabsorption  cannot give information about the Spin-orbit Locking structure. Another way of investigating such structure by exciting Scissors Modes will
be presented in a separate paper\cite{Hata1}. Using Eq.(\ref{phi}) we find the absorption cross-section 
\be
\sigma= {9 \pi^2 \over 4} {\alpha \, \hbar \over m^2c^2} \, {\sqrt {{\mathcal I} C}}\,.
\ee

We can repeat the above calculation with the wave functions appropriate to atoms with axial symmetry~\cite{Hata}. In this case the Scissors states are
labeled by the component of the angular momentum along the cell anisotropy axis, $m=\pm1$, and we find 
\be
\sum_{m=1,-1} \sum_{\lambda} | M_i^{(\lambda)} |^2 = { 9 \over 8 } k^2  \left( {\hat {\bf k}} \cdot {\hat {\bf a}} \right)^2{\sqrt {{\mathcal I} C}}\,,
\ee
where $ {\hat {\bf k}}$, $ {\hat {\bf a}}$ are the unit vectors in the direction of the photon momentum and the anisotropy axis respectively.
The cross-section for photon absorption by crystals with axially symmetric atoms is 
\be
\sigma= {9 \pi^2 \over 2} {\alpha \, \hbar \over m^2c^2}  {\sqrt {{\mathcal I} C}} \left( {\hat {\bf k}} \cdot {\hat {\bf a}} \right)^2 \,.
\ee
It gets its maximum when the photon momentum is parallel to the anisotropy axis
of the cell, and vanishes when it is perpendicular to it. This dichroism is due to the fact that the angular momentum of the photon has the direction of its momentum.
Since the atom can rotate only around the anisotropy axis of the cell, it can absorb the photon angular momentum only in this direction.

\section{Moment of inertia and restoring force constant}

In this Section we make a few considerations about how to make the estimate  of moment of inertia and restoring force constant a bit
more realistic remaining in the framework of a semiclassical model. 

In the case of atoms with axial symmetry and ionic bond an analytic  evaluation of these parameters  was possible 
in the framework of the present semiclassical model under the assumption that the precessing atom can be treated as a rigid body with constant
charge density. Some improvement can easily be obtained by replacing the constant density by some theoretical or phenomenological
function $ \rho({\bf r})$.
 The expressions of moment of inertia and restoring force constant can no longer be given in closed form, but can easily be evaluated. For instance  the potential energy in the crystalline field $V_c$ of a charge density $\rho$ rotated through the
angle $\phi$ around the $i$-axis is given by the expression $\int d {\bf r } V_c ({\bf r}) e^{ -i \phi L_i} \rho({\bf r})$, where $L_i$ is the i-th component of the angular momentum,
so that 
\be
C \approx  \int d {\bf r } V_c ({\bf r})  L_i^2  \, \rho({\bf r}) \,.
\ee
More important changes can be devised thinking about a microscopic calculation based on a Random Phase Approximation. In such a framework one can neglect electrons of the atomic core weekly coupled to the electrons responsible for deformation, which are the ones
participating to the collective motion. This feature is already  present in our evaluation of the restoring force constant (in which only the electrons external to the maximal sphere inscribed in the charge profile contribute), but not in the evaluation of the moment of inertia. It can be interesting to compare with the corresponding situation in Nuclear Physics. 
  For deformed atomic nuclei it is known that the actual value of the
moment of inertia is smaller than that of a rigid body. A semiquantitative estimate in this case can be obtained by neglecting the contribution of the nucleons
internal to the largest sphere inscribed in the nucleus as shown in Fig.2, getting
 \be
{\mathcal I}_{\mbox{effective}} =\frac{r_M^2 - r_m^2}{r_M^2 + r_m^2} \,\, {\mathcal I}_{\mbox{classic}}
\approx 3 \, {\mathcal I}_{\mbox{classic}} \, \delta \label{I}
\ee
where $ {\mathcal I}_{\mbox{classic}}$ is the classical value of the moment of inertia of an ellipsoid whose major and smaller axes, 
 $r_M,r_m$, are parametrized according to
$
r_M= r_0 ( 1+ 2  \delta)\,,   \,\,\, r_m= ( 1-  \delta) .
$
This has a suggestive interpretation in
the fact that since deformed atomic nuclei are superfluid, their external part can rotate with respect to the internal spherical part
without friction~\cite{Loiu1}. But the actual justification is provided by the agreement with microscopic calculations~\cite{Ende}.
In the case of deformed atoms we cannot invoke 
the same argument, but we can ignore the electrons of the core which are weekly coupled to the electrons responsible for deformation. What is the fraction of these electrons 
is a dynamical problem. Probably the actual value of the moment of inertia lies between the classical and the effective ones.

The procedure just outlined might be regarded as a way to implement at the semiclassical level the quantum
mechanical requirement that a spherical object cannot have a rotational energy: if the moment of inertia according to Eq.~(\ref{I}) is proportional to $|\delta|$,
the rotational energy of the ellipsoid becomes infinitely large for $|\delta| \rightarrow 0$ and therefore rotational states do not exist in this limit.

 Using the expression of the restoring force constant determined in ~\cite{Hata}, and the effective value of the moment of inertia the square of the Scissors Mode frequency becomes
\be
\omega^2  \approx 486 \, Z_l \, { e^2 \over m R_0^3}  \,\delta
\ee
where $Z_l$ is the number of ligands and $R_0$ the distance of the ligands from the center of the cell. We see that the excitation energy does not depend on the size or the number of electrons of the atom at the center of the cell, but only on its deformation and the number of ligands, their charge and distance from the cell center. The replacement of $ {\mathcal I}_{\mbox{classic}}$
by ${\mathcal I}_{\mbox{effective}}$ increases the energy of the Scissors Mode which in the case of LaMnO$_3$ becomes approximately 9 eV.

The above results can immediately be extended to some of the crystals with cubic symmetry. They are the crystals whose atoms
have a charge profile which can be approximated by superimposing three ellipsoids with the same center and the axes in the $\xi_i$ directions
 (see for instance Ref. \cite{Masl}). Therefore both moment of inertia and restoring force constants have twice the values for an ellipsoidal atom.

We distinguish  two cases. The first one is epitomized by transition metal oxides of 
cubic symmetry (like NiO), where the valence shell contributing to the mode is 
the outermost shell (3d orbitals in the first transition metal series). In this case the 
Crystal Field  is of the order of $1 \sim 2$ eV, indicating that a hybridization 
model approach is very appropriate. The second one is exemplified by rare earth 
oxides of the same symmetry (like EuO), where the active shell (4f shell) is 
shielded by occupied external shells (e.g 5s and 5p). In this latter case the 
effective Crystal Field is very small (roughly one tenth of that of the transition metals), which also 
indicates that a Crystal Field model might be appropriate. However this does not mean that 
the 4f electrons are free to rotate (i.e. they are not core states) since 
there is experimental evidence [12] that these orbitals are somehow bound to the 
lattice, though by an hybridization energy which is one order of magnitude less 
than in the case of transition metals. Therefore in both cases, in order to calculate the 
restoring force, we can adopt a hybridization model. In this model, it is known 
that the hybridization energy of a bond is given by the gain in electrostatic 
energy of the electron density accumulated along the bond, in order to take 
advantage of the attraction of both the central atom atomic core and the ligand 
core. In formulas:
$$
{\rm E}_{Hyb} = \int d{\bf r} 
\left (\frac{Z_c}{|{\bf r} - R_c|} + \frac{Z_l}{|{\bf r} - R_l|} \right) 
                \rho({\bf r})
$$
with obvious meaning of the symbols. 
Therefore, the restoring force constant is given by 
$$
C = \int d{\bf r} 
\left( \frac{Z_c}{|{\bf r} - R_c|} + \frac{Z_l}{|{\bf r} - R_l|} \right) 
                L_i^2 \rho({\bf r})
$$
where $L_i$ is the appropriate component of the angular moment.

 \begin{figure}[htbp]
     \resizebox{100mm}{!}{\includegraphics{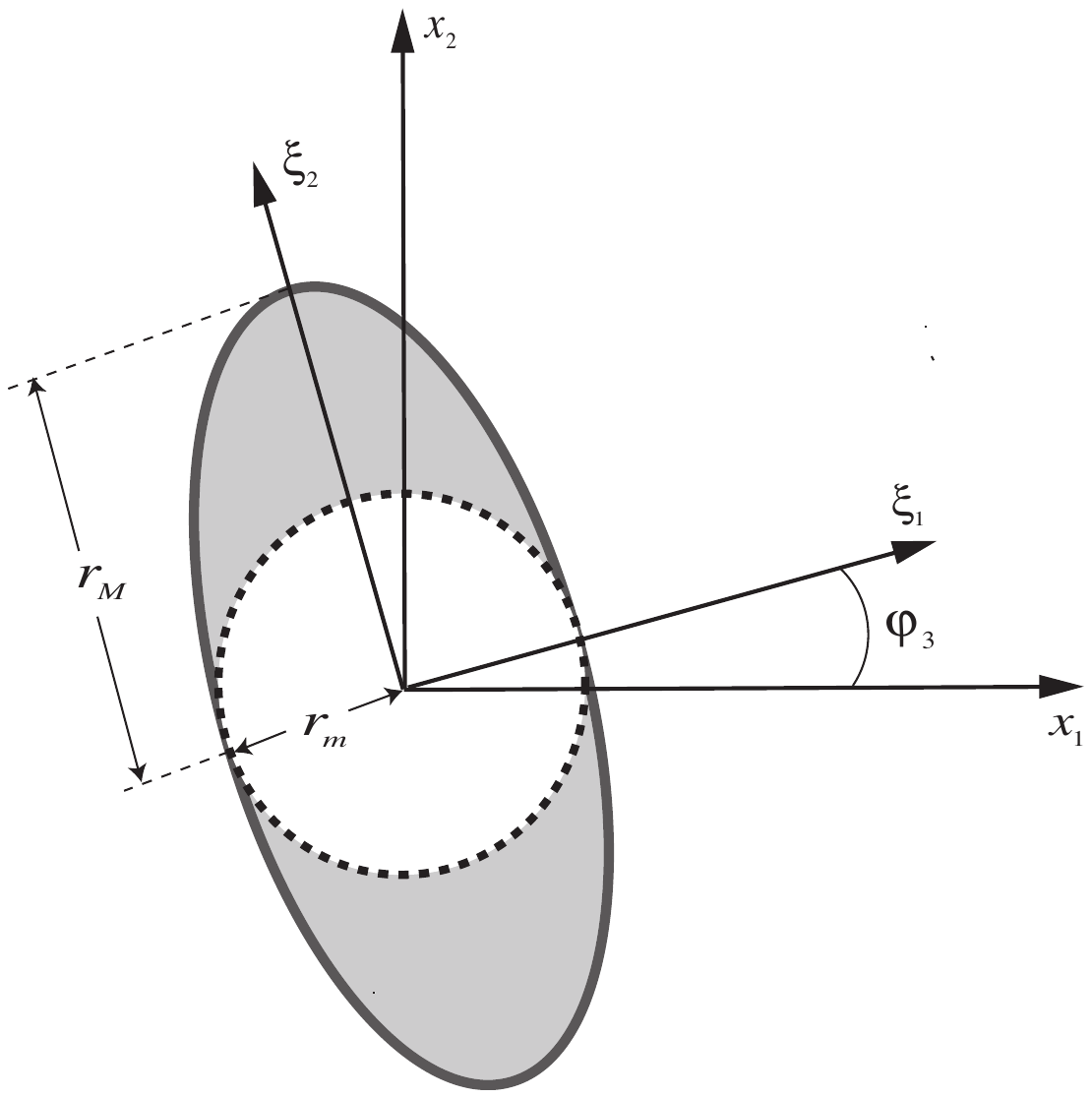}}
         \caption{Atom with axial symmetry. Only the charge external to the sphere of radius $r_m$ contributes to both the restoring
         force constant and the moment of inertia. }
\end{figure}

\section{Conclusion}

In a previous paper we suggested, in the framework of a semiclassical model, that axially symmetric atoms in crystals
with ionic bond should exhibit a collective excitation which should be interpreted as a precession around  the anisotropy axis of their cells.
Such excitation has a magnetic character and a clear cut signature due to the dichroism of the photoabsorption cross-section.

In the present work we studied how the above picture must be modified in crystals with cubic symmetry. This problem is
interesting per se and because of its connection with compounds whose magnetic anisotropy is at the origin of many technological applications.
Among such compounds some might have a Spin-orbit Locking structure which, as already anticipated, might be investigated by excitations of the Scissors Modes\cite{Hata1}.

The main result is that a collective mode of magnetic character can be expected also in such crystals, with a photoabsorption cross-section of
the same order of magnitude as that for crystals with axially symmetric atoms. This cross-section, however, 
does not depend on the angles between the direction of the photon and the axes of the cell.

The evaluation of the parameters of the model is much more difficult in the presence of a covalent bond, in which case the very existence of
the Scissors Mode becomes uncertain. Indeed its basic assumption is that the moving atoms should not change their shape. If the shape
changes, intrinsic motions of electrons will couple to the collective motion, with the possibility of a significant fragmentation.
In fact while in Bose-Einstein condensates the Scissors Mode fully shows
its collective character \cite{Mara}, in atomic nuclei it is split
into two or more close levels. If fragmentation is too high, the magnetic strength is distributed on a quasi continuum of levels
and the collectivity of the state is altogether disrupted. The question therefore is to which extent semiclassical states of the type considered
can exist in a fully quantum framework as a covalent bond. This and the problem of a microscopic evaluation of the moment of inertia are issues 
of conceptual interest.


\subsection *{Acknowledgment}

We are grateful to J. Chaboy and C. Natoli for invaluable comments about crystals with cubic symmetry  and to P. Kr\"uger for a useful correspondence about atoms in the Rare Earth region.

 \end{document}